\documentclass[twocolumn,showpacs,preprintnumbers,amsmath,amssymb,prd]{revtex4}
\usepackage{graphicx}
\usepackage{dcolumn}
\usepackage{bm}
\usepackage{url}

\newcommand{\bq}    {\begin{equation}}
\newcommand{\eq}    {\end{equation}}
\newcommand{\bqr} {\begin{eqnarray}}
\newcommand{\eqr} {\end{eqnarray}}

\begin{document}
\title{Generation of strong magnetic fields by $r$-modes in millisecond
  accreting neutron stars: induced deformations and gravitational wave emission}

\author{Carmine Cuofano$^1$}
\email{cuofano@fe.infn.it}
\author{Simone Dall'Osso$^2$}
\author{Alessandro Drago$^1$}
\author{Luigi Stella$^3$}

\affiliation{$^1$~Dipartimento di Fisica, Universit\'a di Ferrara 
and INFN sezione di Ferrara, 44100 Ferrara, Italy\\
$^2$~Racah Institute of Physics, The Hebrew University of Jerusalem, Giv'at Ram, 91904, Jerusalem, Israel\\
$^3$~INAF - Osservatorio Astronomico di Roma, Via Frascati 33, 00044 Rome, Italy.}

\begin{abstract}
Differential rotation induced by the $r$-mode instability can
generate very strong toroidal fields in the core of accreting, millisecond
spinning neutron stars. We introduce explicitly the magnetic damping term in
the evolution equations of the $r$-modes and solve them numerically 
in the Newtonian limit, to follow the development and growth 
of the internal magnetic field. We show
that the strength of the latter can reach large values, $B \sim 10^{14} $ G, in
the core of the fastest accreting neutron stars. This is strong enough to
induce a significant quadrupole moment of the neutron star mass
distribution, corresponding to an ellipticity $|\epsilon_{\rm B}| \sim 10^{-8}$. 
If the symmetry axis of the induced
magnetic field is not aligned with the spin axis, the neutron star radiates
gravitational waves. We suggest that this mechanism may explain the upper limit of the spin frequencies
observed in accreting neutron stars in Low Mass X-Ray Binaries. 
We discuss the relevance of our results for the search of gravitational waves.
\end{abstract}

\pacs{04.40.Dg, 04.30.Tv, 04.30.Db, 97.60.Jd}

\maketitle

\noindent 
\section{Introduction}
The $r$-mode instability plays an important role in the physics of millisecond neutron
stars NSs. It excites the emission of gravitational waves
(GWs), which carry away spin angular momentum causing the star to spin
down. It also gives rise to large scale mass drifts, particularly in the azimuthal 
direction, and to differential rotation~\cite{Rezzolla2000ApJ,Rezzolla:2001di,Rezzolla:2001dh,Sa:2004gn,Sa:2006hn}.
Differential rotation in turn produces very strong toroidal magnetic fields inside 
the star and these fields damp the instability by tapping the 
energy of the modes.
This mechanism has been investigated in the case of rapidly rotating,
isolated, newly born neutron stars in Refs.~\cite{Rezzolla2000ApJ,Rezzolla:2001di,Rezzolla:2001dh,Abbassi:2012MNRAS}
and in the case of accreting millisecond neutron and quark stars in Refs.~\cite{Cuofano:2009yg,Bonanno:2011vu}.
Magnetic fields deform the star and if the magnetic axis is not
aligned with the rotation axis the NS undergoes free body precession. The
deformation induced by the strong toroidal field is such that the
symmetry axis of the precessing NS drifts on a timescale determined 
by its internal viscosity, eventually becoming an orthogonal rotator~\cite{Cutler:2002nw}.
This is an optimal configuration for efficient GW emission, thus enhancing 
angular momentum losses from the NS.  
\\ 
In this work we resume the analysis in Ref.~\cite{Cuofano:2009yg} about the growth 
of the internal magnetic fields in accreting millisecond neutron stars induced by
$r$-modes. We account for the back-reaction of the generated
magnetic field on the mode amplitude, by introducing the magnetic 
damping rate into the evolution equations. 
\\
We find that the GW emission due to the magnetic
deformation grows secularly in accreting NSs, effectively limiting the
growth of their spin frequency. In particular, this mechanism could 
naturally explain the observed cut-off above $730$~Hz in the distribution of 
spin frequencies of the faster accreting neutron stars in Low Mass X-Ray
Binaries (LMXBs)~\cite{Chakrabarty:2003kt,Chakrabarty:2004tp}.
\\
Finally, we estimate the strain amplitude of the GW signal 
emitted from LMXBs in the local group of galaxies to establish their detectability 
with advanced LIGO and Virgo interferometers. \\
The paper is organized as follows: in Sec. II we write the r-mode equations
taking into account also the magnetic damping rate; in Sec. III we discuss
the magnetic deformation of the star and the associated gravitational-waves emission;
in Sec IV we derive the expression of the magnetic damping rate;
in Sec. V we discuss numerical solutions of the equations;
in Sec. VI we deal with the evolutionary scenarios for accreting neutron stars;
in Sec. VII we focus on the relevance of our results for the search of gravitational waves. 
Finally in Sec. VIII we summarize our conclusions.

\section{R-mode equations}
\begin{figure}
\begin{center}
\includegraphics[height=7cm, width=9cm]{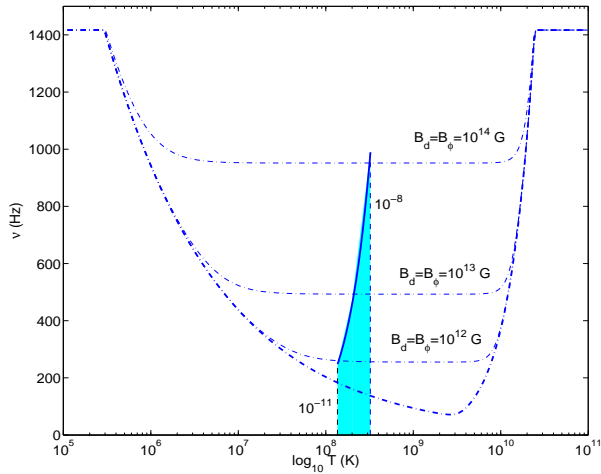}
\end{center}
\caption{\label{fig1} 
R-mode instability regions for NSs with strong internal fields. 
The $r$-mode instability window without internal magnetic field corresponds to the thick dotted-dashed line.
The vertical dashed lines indicate the internal equilibrium
temperatures of accreting neutron stars in LMXBs for, respectively, a
maximum mass accretion rate $\dot{M}=10^{-8}\mbox{M}_\odot\mbox{yr}^{-1}$ (right) and
a minimum value of $\dot{M}=10^{-11}\mbox{M}_\odot\mbox{yr}^{-1}$ (left).
The instability window with a stable configuration
of mixed poloidal-toroidal fields in the inner core, obtained from
the condition $F_g^r=F_{m_i}+F_\nu$, is indicated by thin dotted-dashed lines. 
Three different values of the internal magnetic field are considered and we assume
that poloidal and toroidal components have similar strengths.
The light blue area indicates the region accessible to stable accreting NSs (see Sec.~\ref{evol}). 
The thick solid line indicates the maximum 
spin frequencies when the emission of gravitational waves
due to the star's deformation is taken into account ($\bar{B}_{\phi}\sim 10^{14}$~G, $|\epsilon_B|= 10^{-8}$).}
\end{figure}

We derive here the equations that describe the evolution 
of r-modes in a NS core with a pre-existing poloidal magnetic field, $B_p$, 
and the associated generation of an internal azimuthal field $B_\phi$. 
Following Ref.~\cite{Wagoner:2002vr}, the total angular momentum $J$ of a 
star can be decomposed into an equilibrium angular momentum $J_{*}$ and a 
canonical angular momentum $J_{c}$ proportional to the $r$-mode perturbation:
\begin{equation}
J=J_*(M,\Omega)+(1-K_j)J_c, \,\,\,\,\,\,\,\,\,\,\,\,\,  J_c=-K_c\alpha^2J_*
\label{eq1}
\end{equation}
where $K_{(j,c)}$ are dimensionless constants and
$J_{*}\cong I_{*}\Omega$. \\
The canonical angular momentum obeys the following equation~\cite{Cuofano:2009yg,Friedman:1978hf}: 
\begin{eqnarray}
dJ_c/dt =&& 2J_c\{F_{g}^{r}(M,\Omega) \nonumber \\
&& -[F_v(M,\Omega,T_v)+F_{m_i}(M,\Omega,B)]\}
\label{eq2}
\end{eqnarray}
where $F_g^r$ is the gravitational radiation growth rate of
the $r$-mode, $F_v=F_s+F_b$ is the sum of the shear and bulk viscous damping
rate and $F_{m_i}$ is the damping rate associated with the
generation of an internal magnetic field. Finally, $T_v(t)$ is a 
spatially averaged temperature. \\
The total angular momentum satisfies the equation:
\begin{equation}
dJ/dt=2J_c F_g^r+\dot{J}_a(t)-I_{*}\Omega F_{m_e}
\label{eq3}
\end{equation}
where $\dot{J}_a$ is the rate of variation of angular momentum due to mass
accretion (we assume $\dot{J}_a=\dot{M}(GMR)^{1/2}$, see
Ref.~\cite{Andersson:2001ev}) and $F_{m_e}$ is the magnetic braking
rate associated to the external poloidal magnetic field. In this work
we do not consider additional torques such as,
e.g. the interaction between the magnetic field and the accretion
disk.  Combining Eqs.~(\ref{eq2}) and (\ref{eq3}) we obtain the
evolution equations of the r-mode amplitude $\alpha$ and of the
angular velocity of the star $\Omega$:
\begin{eqnarray}
\frac{d\alpha}{dt} =&& \alpha(F_g^r-F_v-F_{m_i}) \nonumber \\
&& +\alpha[K_jF_g^r+(1-K_j)(F_v+F_{m_i})]K_c\alpha^2 \nonumber \\
&& -\frac{\alpha\dot{M}}{2\tilde{I}\Omega}
\left(\frac{G}{MR^3}\right)^\frac{1}{2}+\frac{\alpha F_{m_e}}{2} 
\label{eq4} \\
\frac{d\Omega}{dt} =&& -2K_c\Omega\alpha^{2}
[K_jF_g^r+(1-K_j)(F_v+F_{m_i})] \nonumber \\
&& -\frac{\dot{M}\Omega}{M} +\frac{\dot{M}}{\tilde{I}}
\left(\frac{G}{MR^3}\right)^\frac{1}{2}
-\Omega F_{m_e} \;
\label{eq5}
\end{eqnarray}
where $I_*=\tilde{I}MR^2$ with $\tilde{I}=0.261$ for an n=1 polytrope
and $K_c=9.4\times 10^{-2}$, see Ref.~\cite{Owen:1998xg}. Our results
turn out to be rather insensitive to the exact value of $K_j \sim 1$ (see
Ref.~\cite{Wagoner:2002vr}).
\\ 
The classical $r$-mode equations are simply recovered from the above by neglecting the
  magnetic damping term, $F_{m_i}$. In this case, it is readily seen that
  the $r$-modes will grow only if the condition $F^{r}_{\rm g} > F_{\rm v}$
  is satisfied, while viscosity keeps the NS stable against $r$-modes
  if this condition is reversed. This simple argument defines the classical
  instability window for $r$-modes, which is drawn in Fig.~\ref{fig1} as the 
thick  dot-dashed curve.

Well inside the classical instability window, where viscosity is negligible, 
a new type of equilibrium can be reached. 
Eq.~(\ref{eq4}) and Eq.~(\ref{eq5}) indeed imply that the
growth of $r$-modes can be quenched by the magnetic damping term, 
$F_{\rm m_i}$, when the condition $F_g^r \leq F_{m_i}$ holds. 
Since $F_{\rm g}^r=F_{\rm g}^{\rm r} (\Omega)$ and $F_{m_i}=F_{m_i}(\Omega,B)$, the equilibrium
  condition $F_{\rm g}^r=F_{m_i}$ implies a relation $\Omega=\Omega(B)$, which can
  be written as:
\begin{eqnarray}
 \nu_{cr}&\approx& 68 \,\,B_{d,8}^{\,1/7} \,\,\bar{B}_{\phi,12}^{\,1/7} \,\,M^{-2/7}_{1.4} \,\,R_{10}^{-3/7}
\label{ncr}
\end{eqnarray} 
where $B_{d,8}=(B_d/10^8 \,\mbox{G})$, $\bar{B}_{\phi,12}=(\bar{B}_\phi/10^{12} \,\mbox{G})$,
$M_{1.4}=(M/1.4 \,M_\odot)$, $R_{10}=(R/10 \, \mbox{km})$. 
\\
Here, and in the following, $\nu$ indicates the NS spin frequency and
 $\bar{B}_\phi$ is an energy averaged, azimuthal magnetic field induced by the
$r$-modes.

For a given combination of $({\rm B_{\rm d}}, \bar{B}_\phi)$, Eq.~(\ref{ncr}) represents 
the critical frequency above which the star is $r$-mode unstable even against
the effect of the magnetic damping term. For a given value of the spin
  frequency, on the other hand, Eq.~(\ref{ncr}) tells us the minimal combination
  of field strengths required to damp the $r$-modes. With the value of
    $B_{d}$ fixed, $r$-modes are excited as long as $\bar{B}_\phi$ is lower than required by
  Eq.~\ref{ncr}. They continue to generate further azimuthal field, until Eq.~(\ref{ncr})
is satisfied and the NS is eventually stabilized with respect to r-modes.

\section{Magnetic deformation and GW emission}
\label{sezione3}
We assume a NS with an internal poloidal field which is described by
Ferraro's solution~\cite{Cuofano:2009yg,Ferraro1954ApJ} 
\begin{eqnarray}
&&\mbox{\textbf{B}}^{in}(t=0)=\\ \nonumber
&& B_d \, \left[\left(-3\frac{r^2}{R^2}+5\right)
\texttt{cos} \, \theta \,\mbox{\textbf{e}}_r+\left(6\frac{r^2}{R^2}-5\right)\, \texttt{sin} \,
\theta \, \mbox{\textbf{e}}_\theta\right] \, .
\label{eq1Fe}
\end{eqnarray}
where $B_d$ is the surface field strength at the stellar equator. 
This is matched to an exterior dipolar field aligned with the star spin axis
\begin{eqnarray}
\mbox{\textbf{B}}^{ext}=B_d \frac{R^3}{r^3}(2
\, \texttt{cos} \, \theta \, \mbox{\textbf{e}}_r+ \, \texttt{sin} \,
\theta \, \mbox{\textbf{e}}_\theta)
\label{eq1M}
\end{eqnarray}
As for the internal structure of the NS, it is  widely
  accepted that protons will be superconducting at temperatures 
$T \lesssim 10^9$~K~\cite{Baldo:2007,Wasserman:2003}, at least in a fraction of the NS core. 
To maintain focus on the salient properties of our scenario, we carry out 
here a detailed study of the growth of r-modes and of the azimuthal field in a
normal fluid core. In Section~\ref{supercond} we consider the likely occurrence of a shell of
superconducting protons in the outer core and assess its impact on our conclusions.
\\
Once modes are excited at an initial time $t_0$, they induce azimuthal
drift motions of fluid parcels in the NS core. Following Ref.~\cite{Rezzolla:2001di}, 
the total angular displacement from the onset of the instability up to time $t$ reads
\begin{eqnarray}
\label{eq.delta-xphi}
\Delta \tilde{x}^{\phi}(r,t) = \frac{2}{3}\left(\frac{r}{R}\right)
k_2(\theta)\int_{t_0}^t\alpha^2(t')\Omega(t')dt'+\mathcal{O}(\alpha^3) \,\,\,\,\,\,\,\,\,
\end{eqnarray}
where $k_2(\theta)\equiv (1/2)^7(5!/\pi)(\texttt{sin}^2\theta-2\texttt{cos}^2\theta)$.
Magnetic field lines are twisted and stretched by the shearing motions
and a new field component, in the azimuthal direction, is generated accordingly. 
The relation between the new and the original magnetic field inside the star 
in the Lagrangian approach is~\cite{Rezzolla:2001di}:
\begin{eqnarray}
\frac{B^j}{\rho}(\tilde{\mbox{\textbf{x}}},t)=\frac{B^k}{\rho}(\mbox{\textbf{x}},t_0)
\frac{\partial\tilde{x_j}(t)}{\partial x^k(t_0)}.
\end{eqnarray}
This equation implies that the radial dependence of the initial and 
final magnetic field is the same.
Integrating over time the induction equation in the Eulerian approach one gets~\cite{Rezzolla:2001dh}:
\begin{eqnarray}
\delta B^{\theta} &\simeq& \delta B^r \simeq 0  \label{eq2Ma}\\
\delta B^{\phi} (r,t) &\simeq& B_0^{\theta} \int \dot{\phi}(t') dt' 
\simeq B_0^{\theta} \Delta \tilde{x}^{\phi}(r,t)
\label{eq2M}
\end{eqnarray}
where $B^{\phi}$ is the toroidal component generated by the $r$-modes.
\\
A toroidal field $B^{\phi}$ deforms a neutron star into a prolate ellipsoid with ellipticity
\begin{eqnarray}
\epsilon_B\equiv\frac{I_{zz}-I_{xx}}{I_{zz}}
\end{eqnarray}
where $I_{jk}=\int_V\rho(r)(r^2\delta_{jk}-x_jx_k)dV$ is the inertia tensor.
For a neutron star with a normal core, this gives
\begin{eqnarray}
 \epsilon_B = - k_{\epsilon}\times 10^{-12} ~ R_{10}^4 ~ M_{1.4}^{-2} ~ \bar{B}_{\phi,12}^2
\label{eccB}
\end{eqnarray}
where $k_{\epsilon}\simeq[0.5\mbox{--}5]$ is a parameter depending on the
configuration of the internal fields \cite{Cutler:2002nw,Haskell:2007bh,Mastrano:2011tf}. 
\\
In general, we expect the symmetry axis of the magnetic deformation
not to be perfectly aligned with the rotation axis. In this
situation, $J_*$ precesses around the symmetry axis with a period
$P_{\rm prec}\simeq P/|\epsilon_B|$ \cite{Stella:2005yz, Dal07, Dal09}, $P$ being the spin period of the star.
\\
Due to the neutron star rotation, the relaxed state of its crust also has an
  intrinsic \textit{oblate} deformation. This corresponds to the shape it
  would retain if the NS spin were stopped completely, without the crust
  cracking  or re-adjusting in anyway. The ellipticity associated to this crustal
 reference deformation can be written as~\cite{Cutler:2002np,Cutler:2002nw,Zdunik:2008tc}:

\begin{equation}
\epsilon_{\rm c} \approx k_c\times 10^{-8}\left(\frac{\nu_{\rm s}}{\rm kHz}\right)^2\,.
\label{eccCR}
\end{equation}
where $k_c\simeq[1\mbox{--}3]$ for an accreted crust~\cite{Zdunik:2008tc}.
As long as this is dominant, the NS effective mass distribution will 
be that of an oblate ellipsoid. In this case, the spin and symmetry axis
of the freely precessing NS will tend to align. However, as the internal
toroidal field grows, the ensuing magnetic deformation will at some point overcome the maximal
oblateness that the crust can sustain (Eq.~(\ref{eccCR})). The NS
first enters the r-mode instability window when its spin rate is 
$\nu_{\rm s,i} \gtrsim 150$ Hz. Therefore, the magnetic deformation will first
  become dominant when
\begin{eqnarray}
|\epsilon_{\rm B}| & \gtrsim & 2 \times 10^{-10}~ k_c\left(\frac{\nu_{\rm
    s,i}}{150~\rm Hz}\right)^2 \equiv \epsilon_{B,min}~~~{\rm ~or} \nonumber \\
\bar{B}_\phi & \gtrsim & 1.5\times 10^{13}\, \mbox{G}\left(\frac{k_c}{k_\epsilon}\right)^{1/2}
\left(\frac{\nu_{\rm s,i}}{150\mbox{~Hz}}\right)\,M_{1.4}\,R^{-2}_{10}~.~~~~
\label{eq.Bcr}
\end{eqnarray}
Beyond this point, the total NS ellipticity becomes dominated by the
 magnetically-induced, prolate deformation. The freely precessing NS now becomes secularly 
unstable. In the presence of a finite viscosity of its interior, the wobble
angle between the angular momentum $J_*$ and the symmetry (magnetic)
axis grows until the two become orthogonal. This occurs on a dissipation
(viscous) timescale,  $\tau_{\rm v}$, which can be expressed as~\cite{Cutler:2002nw}
\begin{equation}
 \tau_{\rm v} = n P_{\rm prec} \simeq 10^{11}\times\left(\frac{n}{10^5}\right)\left(\frac{100~\rm{Hz}}{\nu_s}\right)
\left(\frac{10^{-8}}{|\epsilon_B|}\right) \,\, \mbox{s}~~.
\label{eq.Tdis}
\end{equation}
The parameter $n$ measures the dissipation
timescale in units of the free precession cycle. Below, we 
determine quantitatively its role in promoting the orthogonalization of 
the NS symmetry and rotation axes. Given the significant theoretical 
uncertainties on the value of $n$, we normalize it to an educated
guess~\cite{Cutler:2002nw} but will consider it effectively as a free parameter.

An orthogonal rotator with a time-varying quadrupole moment, 
$Q_{22}\approx I_*\epsilon_{\rm B}$, emits gravitational-waves at a rate~\cite{Shapiro83}
\begin{eqnarray}
 \dot{E}_{\rm gw}^B=-\frac{32}{5}\frac{G(I_*\epsilon_B)^2}{c^5}\Omega^6\,\,.
\label{eq.EgB}
\end{eqnarray}
which produces spin down at the rate
$F_{\rm g}^{\rm B}\equiv -(\dot{E}_{\rm gw}^{\rm B}/2E)$
\begin{equation}
 F_{g}^B \simeq \frac{1}{5\times10^{14}}\, \tilde{I_1}\, M_{1.4}
~R_{10}^{2}\, \left(\frac{\epsilon_{\rm B}}{10^{-8}}\right)^{2}\, P_{-3}^{-4} ~ \mbox{s}^{-1} ~
\label{eq.FgB}
\end{equation}
where $\tilde{I}_{1}=\tilde{I}/0.261$.
\\
Therefore, if the orthogonality condition is satisfied, the NS looses
significant spin angular momentum to GWs also through the 
induced magnetic deformation. This is accounted for by introducing the
appropriate GW radiation term, $F^{\rm B}_{\rm g}$, in the Eq.~(\ref{eq3}), which now becomes:
\begin{equation}
 dJ/dt=2J_c F_g^r+\dot{J}_a(t)-I_*\Omega(F_g^B+F_{m_e})\, \, .
\label{torque2}
\end{equation}
Eq.~(\ref{eq4}) and Eq.~(\ref{eq5}) for the $r$-mode amplitude and the NS spin frequency
are modified accordingly:
\begin{eqnarray}
\frac{d\alpha}{dt} =&& \alpha(F_g^r-F_v-F_{m_i}) \nonumber \\
&& +\alpha[K_jF_g^r+(1-K_j)(F_v+F_{m_i})]K_c\alpha^2 \nonumber \\
&& -\frac{\alpha\dot{M}}{2\tilde{I}\Omega}
\left(\frac{G}{MR^3}\right)^\frac{1}{2}+\frac{\alpha (F_g^B+F_{m_e})}{2} \,, 
\label{eq6} \\
\frac{d\Omega}{dt} =&& -2K_c\Omega\alpha^{2}
[K_jF_g^r+(1-K_j)(F_v+F_{m_i})] \nonumber \\
&& -\frac{\dot{M}\Omega}{M} +\frac{\dot{M}}{\tilde{I}}
\left(\frac{G}{MR^3}\right)^\frac{1}{2}
-\Omega (F_g^B+F_{m_e}) .~~
\label{eq7}
\end{eqnarray}
Eq.~(\ref{eq6}) and Eq.~(\ref{eq7}) now contain two terms for GW emission.
Of these, $F^{B}_{g}$ is given by Eq.~(\ref{eq.FgB}) while 
we adopt for $F^{r}_{g}$ the $r$-mode
gravitational radiation reaction rate due to the
$l=m=2$ current multipole~\cite{Andersson:2000mf} 
\begin{equation}
F_{g}^r = \frac{1}{47} ~ M_{1.4} ~ R_{10}^{4} ~ P^{-6}_{-3} ~~ \mbox{s}^{-1}~~.
\label{eq0A}
\end{equation}

The evolution equations (\ref{eq6},\ref{eq7}) will hold only \textit{after} 
the symmetry axis of the azimuthal field has become orthogonal to the spin
axis. \\
If $t$ measures 
the time since the generated magnetic field exceeds the critical value given by Eq.~(\ref{eq.Bcr}),
the rotation of the magnetic axis will occur when 
$t \geq \tau_{\rm v} (t)$, where $\tau_{\rm v}$ is given by Eq.~(\ref{eq.Tdis}).
Once the condition $t\geq\tau_{\rm v}(t)$ is met and Eq.~(\ref{eq6}) becomes effective, it is 
relevant to know which of the two GW emission terms is dominant. To
see this, we consider the ratio between the GW torque due to r-modes, $\dot{J}_g^r=2J_{\rm c}F_g^r$,
and the GW torque due to the magnetic deformation $\dot{J}_g^B=I_*\Omega F_g^B$ (see Eq.~(\ref{torque2}))
\begin{equation}
\frac{\dot{J}_g^r}{\dot{J}_g^B}\approx 2\times 10^{13}\,\alpha^2 \,R_{10}^{2}
\left(\frac{\epsilon_B}{10^{-8}}\right)^{-2}\,P^{-2}_{-3}
\label{eq.ratioGW}
\end{equation} 
The magnetic deformation becomes dominant only when the $r$-mode
amplitude gets very small, and the internal magnetic field accordingly
large. In particular, the ratio in Eq.~(\ref{eq.ratioGW}) becomes smaller than unity
once $\alpha$ has decreased below the critical value:
\begin{equation}
\alpha_{\rm cr} \lesssim 2 \times 10^{-7} \,R^{-1}_{10}\,\left(\frac{|\epsilon_{\rm B}|}{10^{-8}}\right)
\,P_{-3} ~~ .
\label{eq.alphamax}
\end{equation} 
A numerical solution of Eqs.~(\ref{eq6},\ref{eq7}) is required to verify
when the condition given by Eq.~(\ref{eq.alphamax}) is satisfied during the system's evolution. 
Before calculating the evolution of $\alpha$ in detail, we note that 
a new asymptotic equilibrium could take place in this situation, with
the material torque being balanced by the GW torque due to the magnetic deformation,  
$F_{\rm g}^{\rm B}$. This new equilibrium condition,
 $\dot{J}_a = \dot{J}_g^B$, leads to a limit spin frequency 
\begin{equation}
 \nu_{max}\approx 990 ~\tilde{I}_{1}^{-2/5} M_{1.4}^{-3/10} R_{10}^{-7/10}
\,\left(\frac{\epsilon_{\rm B}}{10^{-8}}\right)^{-2/5}\,\dot{M}_{-8}^{1/5} ~\mbox{Hz}
\label{nmax1}
\end{equation}
where $\dot{M}_{-8}=\dot{M}/(10^{-8} \,\rm M_{\odot} \, \rm{yr}^{-1})$.

\section{Magnetic damping rate}

Following Ref.~\cite{Rezzolla:2001dh}, we make use of Eq.~(\ref{eq.delta-xphi}) and Eq.~(\ref{eq2M})
to derive the expression for the magnetic damping term. Let us first write 
the variation of the magnetic energy for a NS with a normal fluid core
\begin{eqnarray}
 \delta E_M \equiv \frac{1}{8\pi}\int_{V_\infty}\delta B^2 dV
\label{varEm}
\end{eqnarray}
where $\delta B^2 = (\delta B_{\hat{p}})^2+(\delta B_{\hat{\phi}})^2\simeq(\delta B_{\hat{\phi}})^2=B_{\hat{\phi}}^2$.
In the last steps we have taken into account the Eqs.~(\ref{eq2Ma},\ref{eq2M}).
The rate of variation of magnetic energy can be obtained from the Eqs.~(\ref{eq.delta-xphi},\ref{eq2M})
and reads
\begin{eqnarray}
 \delta E_M^N &\simeq& \frac{2AB_d^2R^3}{9\pi}\left(\int_0^t\alpha^2(t')\Omega(t')dt'\right)^2 ~.
\label{dEM-dt}
\end{eqnarray}
From this the expression of the magnetic damping rate is derived~\cite{Cuofano:2009yg}:
\begin{eqnarray}
F_{m_i}^N &=& \frac{(dE_M^N/dt)}{\tilde{E}} \nonumber \\ 
&\simeq& \frac{4 A}{9\pi \cdot (8.2\times 10^{-3})}
\frac{B_d^2 R\int^t_0 \alpha^2(t ')\Omega(t ') dt '}{M\Omega}\,\,\,\,\,\,\,\,\,\,\,
\label{eq8}
\end{eqnarray}
where $\tilde{E}\simeq 8.2\times 10^{-3} \alpha^2 M \Omega^2 R^2$ is the energy of the $r$-mode~\cite{Rezzolla:2001dh}
and $A\simeq 0.99$~\cite{Cuofano:2009yg}.
We make use of Eqs.~(\ref{varEm}) and (\ref{dEM-dt}) to estimate the toroidal magnetic 
field generated by r-modes
\begin{eqnarray}
 \bar{B}_\phi\simeq \left(\frac{4A}{3\pi}\right)^{1/2} B_d \int_0^t\alpha^2(t')\Omega(t')dt' \,.
 \label{BphiEv}
\end{eqnarray}
Note that $F_{\rm m_i}$ is not related to $\alpha$ in a simple way. The
magnetic damping rate depends, like $B_{\hat{\phi}}$, on the integrated
history of the $r$-mode amplitude since the star first entered the instability
window (cfr. Eq.~(\ref{eq2M})).
\\ 
Finally, by using the volume-averaged expression for the azimuthal
field, $\bar{B}_{\phi}$, we can re-write the expression of the magnetic damping rate as
\begin{eqnarray}
F_{m_i}^N \approx \frac{1}{6.67\times10^9}  \,R_{10} \,M_{1.4}^{-1} \,B_{d,8} \,\bar{B}_{\phi,12} \,P_{-3} ~ \mbox{s}^{-1} .~
\label{Fmi}
\end{eqnarray}

\section{Numerical solutions}
In the scenario we are describing, an initially slowly rotating NS is
secularly spun up by mass accretion. When its spin frequency reaches a  few
hundred Hz, the NS enters the classical $r$-modes instability window. As the instability
develops, the evolution of the $r$-mode amplitude becomes coupled to the
growth of an internal, toroidal magnetic field. We now turn to a detailed
numerical calculation of their coupled evolution, in the light of our previous
discussion. This will help us clarify the sequence of events
expected to occur as an accreting NS in a LMXB is secularly spun up by the material torque.
In all our calculations we consider values of the mass accretion rate 
$\dot{M}=(10^{-8}\,\mbox{--}\,10^{-10}) \,\mbox{M}_{\odot}\mbox{yr}^{-1}$. Note that,
in accreting NSs, the mass accretion rate has an 
upper limit $\dot{M}_{\rm Edd}\sim 10^{-8} \mbox{M}_{\odot}\mbox{yr}^{-1} $ and
most LMXBs do not accrete at this rate for a long time. 
Here $\dot{M}_{\rm Edd}$ is the mass accretion rate that produces the Eddington luminosity.

\subsection{Numerical estimates of relevant rates}

We begin by discussing the main physical quantities in the 
evolution equations that were not described in previous sections.
\\
For non--superfluid matter the shear viscosity damping rate reads~\cite{Andersson:2000mf}
\begin{equation}
F_{s} = \frac{1}{6.7\times 10^{7}}
M_{1.4}^{5/4}R_{10}^{-23/4}T^{-2}_{9} ~~ \mbox{s}^{-1}~,
\label{eq0B}
\end{equation}
where $T_9=T/10^9$ K, while the bulk viscosity damping rate is given by~\cite{Owen:1998xg} 
\begin{equation}
F_{b} = \frac{1}{6.99\times 10^8}
\left(\frac{\Omega^2}{\pi G \bar{\rho}}\right)T_9^6 ~ ,
\label{eq0C}
\end{equation}
which we can approximately rewrite as
\begin{equation}
F_{b} = \frac{1}{2.5\times 10^{9}} M^{-1}_{1.4}R_{10}^3
P^{-2}_{-3}T^{6}_{9} \,\,\,\,\, \mbox{s}^{-1}\,.
\label{eq0D}
\end{equation}
However, for the temperatures of interest here
the bulk viscosity damping rate is a few orders of magnitude smaller than the
shear viscosity damping rate and therefore it is negligible.
\\
Notice that viscous damping of the modes depends strongly on temperature, and 
that temperature will in turn be affected by viscous heating. 
It is thus important to describe accurately the global thermal balance of the NS.
We consider three main factors: modified URCA cooling ($\dot{\epsilon}_{\rm u}$), shear viscosity
reheating ($\dot{\epsilon}_{\rm s}$) and accretion heating ($\dot{\epsilon}_{\rm n}$). 
The equation of thermal balance of the star therefore reads:
\begin{equation}
 \frac{d}{dt}\left[\frac{1}{2}C_{v}T \right]= 
-\dot{\epsilon}_u+\dot{\epsilon}_s+\dot{\epsilon}_n \,\, .
\label{eq13}
\end{equation}
Here, C$_{\rm v}$ is the total heat capacity of the NS~\cite{Watts:2001ej}:
\begin{equation}
 C_{\rm v}=1.6\times 10^{39}M_{1.4}^{1/3}T_{9} \,\,\,\,\, \mbox{erg K}^{-1}.
\label{eq12}
\end{equation}
The cooling rate due to the modified URCA reactions, $\dot{\epsilon}_{\rm u}$, 
reads \cite{Shapiro83}
\begin{equation}
 \dot{\epsilon}_{\rm u}=7.5\times 10^{39}M_{1.4}^{2/3}T_{9}^{8} 
\,\,\,\,\,\, \mbox{erg s}^{-1} \,.
\end{equation}
If direct URCA processes can also take place, the NS cooling rate is
expected to become much larger~\cite{Blaschke2004A&A}. A recent phenomenological analysis
of their implications can be found e.g. in Ref.~\cite{Blaschke2006PhRvC}.
We assume, for simplicity, that only modified Urca processes are allowed in the NS core. 
\\
Dissipation of the $r$-mode oscillations by the 
action of shear viscosity will contribute to heating the NS. The
corresponding heating rate, $\dot{\epsilon}_{\rm s}$,
reads~\cite{Andersson:2000mf}
\begin{eqnarray}
 \dot{\epsilon}_{\rm s} &=& 2\alpha^2\Omega^2MR^2\tilde{J}F_s \nonumber\\
 &=& 8.3\times10^{37}\alpha^2\Omega^2\tilde{J}M_{1.4}^{9/4}
 R_{10}^{-15/4}T_{9}^{-2} \,\,\,\,\,\, \mbox{erg s}^{-1}  \,\,\,\,\,\,\,\,
\end{eqnarray}
where $\tilde{J}=1.635\times 10^{-2}$.
\\
The accreting material exerts an extra pressure onto the NS crust causing direct
compressional heating. The compression also triggers pycnonuclear reactions
in the crust, which release further heat locally. The total heating rate,
given by the sum of the two contributions, is~\cite{Brown:1997ji} :
\begin{equation}
 \dot{\epsilon}_n=\frac{\dot{M}}{m_{N}}\times 1.5 \,\, \mbox{MeV} = 4\times 10^{51} 
 \dot{M}_{1.4} \,\,\,\,\, \mbox{erg s}^{-1}
\label{eq10}
\end{equation}
where $m_N$ is the mass of a nucleon and $\dot{M}_{1.4}=\dot{M}/1.4 M_{\odot}$ 
is measured in s$^{-1}$.

\begin{figure}
\begin{center}
\includegraphics[height=6cm, width=9cm]{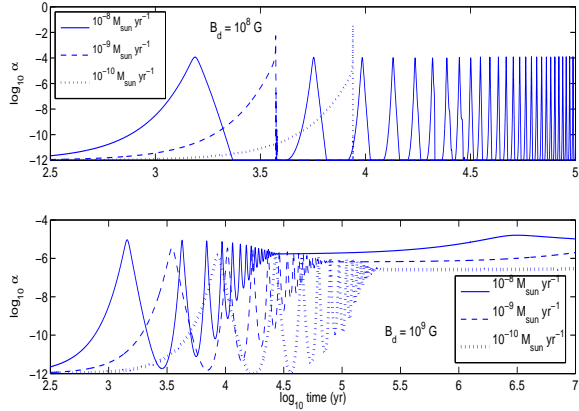}
\end{center}
\caption{\label{fig2} Temporal evolution of the r-mode amplitude $\alpha$ leaving out 
the instabilities of the magnetic fields.
We consider three values of mass accretion rate 
$\dot{M}=(10^{-8},10^{-9},10^{-10})\,\mbox{M}_\odot\,\mbox{yr}^{-1}$ and  two values
of the initial poloidal magnetic field: $B_d=10^8$~G (upper panel) and $B_d=10^9$~G (lower panel).}
\end{figure}

We can now solve self-consistently
Eqs.~(\ref{eq4},\ref{eq5},\ref{eq8},\ref{eq13}), 
to obtain the evolution of the core temperature, $r$-mode
  amplitude and generated internal magnetic field,
  $\bar{B}_\phi$. We show
in Fig.~\ref{fig2} the evolution of the $r$-mode amplitude, $\alpha$, after the star 
first enters the classical $r$-mode instability window. Three different
values of accretion rate $\dot{M}$ and two values of the initial 
poloidal magnetic field $B_d$ are considered. We find that
the maximum values of $\alpha$ are in the range
$\alpha_{max}\sim[10^{-6}-10^{-4}]$. 

\begin{figure}
\begin{center}
\includegraphics[height=6cm, width=9cm]{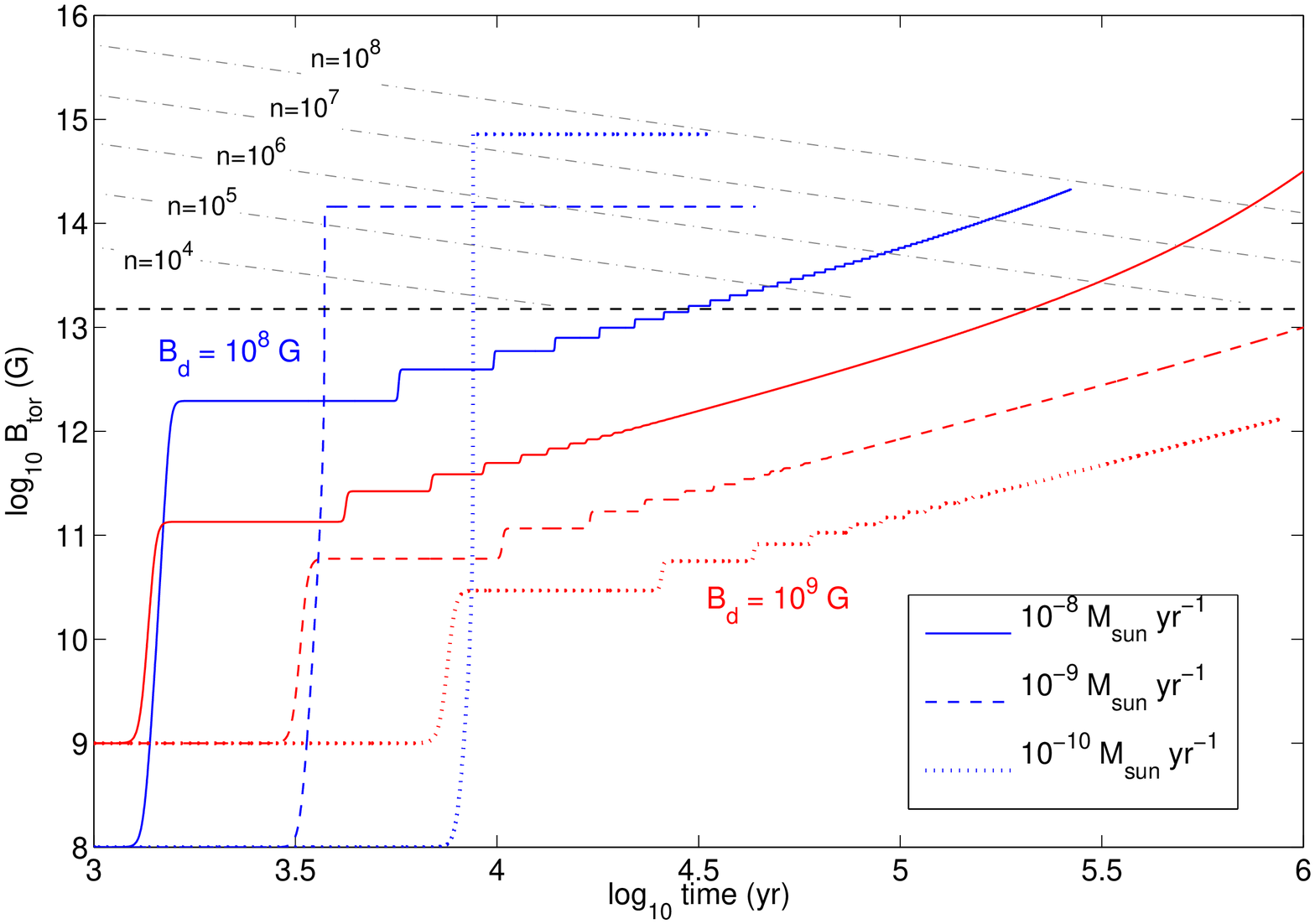}
\end{center}
\caption{\label{fig3} Temporal evolution of the generated toroidal magnetic field
obtained by solving Eqs.~(\ref{eq4},\ref{eq5}).
We considered three values of mass accretion rate
$\dot{M}=(10^{-8},10^{-9},10^{-10})\,\mbox{M}_\odot\,\mbox{yr}^{-1}$
and two values of initial poloidal magnetic field $B_d=(10^8,10^9)$~G.
The black dashed line indicates the minimal strength of the generated toroidal 
magnetic field for which the magnetic deformation becomes dominant (Eq.~(\ref{eq.Bcr})).
The time at which the rotation of the internal field takes place
($t=\tau_{\rm v} (t)$) is shown as a function of time for five values 
of the parameter $n=[10^4\mbox{--}10^8]$ (dashed-dotted gray lines).
Except in cases with large values of initial poloidal fields $B_d\sim 10^9$~G
and small values of mass accretion rate $\dot{M}<10^{-9}\,\mbox{M}_\odot\,\mbox{yr}^{-1}$,
huge magnetic fields $B^{\phi}\sim [10^{13}\mbox{--}10^{14}]$~G are generated in less than $10^6$ years.}
\end{figure}

In Fig.~\ref{fig3} we show the temporal evolution of the generated,
volume averaged toroidal magnetic field. Also plotted are lines along which
the condition $\tau_{\rm v}(t) = t$ is met, for different values of the
parameter $n$. The secular
velocity field associated to the $r$-modes, which is
  predominantly azimuthal, clearly induces very large secular effects. 
In particular, very strong toroidal magnetic fields,
$\bar{B}^{\phi}\sim [10^{13}\mbox{--}10^{14}]$~G, can be produced by the
wrapping of the pre-existing poloidal field lines, in $10^4 - 10^6$ yrs.

The implications of these results for the spin equilibrium condition
(Eq.~(\ref{nmax1})) will be discussed in the next
section. 

\section{The overall evolutionary sequence: a general discussion}
\label{evol}

In the following we describe likely evolutionary scenarios for
accreting ms spinning NSs, based on the numerical solutions and the
physical discussion of the previous
sections. We also consider implications of the possible 
occurrence of the Tayler instability in the strongly twisted internal
field. Finally, the occurrence of proton superconductivity in (at least a
fraction of) the NS core and its impact on our arguments are discussed.

Note that our analysis does not apply to compact stars in which exotic 
matter, like e.g. hyperons or deconfined quarks, is present in the inner core. 
In this type of stars the viscosity and thus the r-mode instability 
window is quite different from that of normal NSs \cite{Drago2008ApJ}. 
Moreover, the evolution of the internal magnetic fields 
and the magnetic properties of the compact star are also expected to be different 
\cite{Bonanno:2011vu,Alford2008RvMP}.
\subsection{Normal Core}
The growth rate of the azimuthal magnetic field once the
accreting NS enters the r-mode instability window depends on the
strength of the initial poloidal field $B_d$, and on the mass accretion rate $\dot{M}$. \\
In the most favorable cases, e.g. for $B_{d,8}=1$ and
$\dot{M}_{-8}=(0.01-0.1)$, the magnetic field can reach a huge strength,
$\bar{B}_\phi\approx 10^{14}$~G, in a few hundred years. 
In other cases, the internal magnetic field evolves on a
much longer timescale up to million years
(see discussion in Ref.~\cite{Cuofano:2009yg}).
\\
As already stated in Sec.~\ref{sezione3}, the NS enters the
$r$-mode instability window at $\nu_{\rm s,i} \gtrsim 150$. 
Hence, a minimal magnetic distortion $|\epsilon_{\rm B, min}| \approx 2\times
10^{-10}$ must be achieved for the magnetic deformation to overcome the crustal
oblateness.  This corresponds to a minimal azimuthal field 
$\bar{B}_{\phi, {\rm min}} \approx 10^{13}$~G. Only once this field strength is reached the 
star becomes secularly unstable. However, the magnetic axis is driven
 orthogonal to the spin axis in a time $\tau_{\rm v}$ (Eq.~(\ref{eq.Tdis})). 
Therefore, the condition $\tau_{\rm v} (t) \leq t$ \textit{must also be satisfied} for tilting
of the magnetic axis to be effective. At the time at which \textit{both}
requirements are met, call it $t_{\rm ort}$, the toroidal magnetic field will
have an intensity $B_{\phi, {\rm ort}}$ and, of course, $\tau_{\rm v} (t_{\rm ort}) \leq t_{\rm ort}$. 
 
After tilting, the original
azimuthal field $B_{\phi, {\rm ort}}$ will have acquired an $r-\theta$
(poloidal) component, whose strength will be comparable to that of the new
$\phi$-component. We assume that 
$\bar{B}_{\rm p,new} = \bar{B}_{\phi,  {\rm new}} \approx (1/\sqrt{2}) \bar{B}_{\phi, {\rm ort}}$, 
although the exact ratio will depend on the details of the
distribution of the internal field.
\\
Note that the generated magnetic fields may directly affect 
the r-mode oscillations when $B \gg 10^{14}$~G \cite{Lee2005,Abbassi:2012MNRAS}. 
We do not include this complicated back-reaction in our calculations. Our
assumption will be justified a posteriori, since the maximum generated 
magnetic fields in our model reach values of order $10^{14}$ G.
\\
In the following we will outline two possible evolutionary paths: which of the two 
will be followed by the star depends on the different combination of the parameters ($n$,$k_\epsilon$,$k_c$),
on the mass accretion rate $\dot{M}$ and of the initial magnetic field $B_d$.
\subsubsection{Evolutionary Path ($\bar{B}_{\phi,ort}\gg \bar{B}_{\phi,min}$)}
The first evolution path we discuss takes place if
the condition $t \gtrsim \tau_{\rm v}$ is met when $B_{\phi, {\rm ort}} \gg B_{\phi,{\rm min}}$. 
In this case a very large deformation of the NS can be obtained. The main dissipation mechanism of these 
magnetic fields in the normal fluid core of neutron stars should be ambipolar diffusion
whose timescale is given by~\cite{Goldreich:1992ap}
\begin{eqnarray}
  t_{amb} \sim 3\times 10^9  \frac{T_8^2 L_5^2}{B_{12}^2}~~\mbox{yr}~,
 \label{t_amb}
\end{eqnarray}
where $L$ is the size of the region embedding the magnetic field and $L_5=L/10^5~\mbox{cm}$.
Note that if $B_{\phi, {\rm ort}}\gg 10^{14}$~G
the generated internal field would decay on a time shorter than a few million years
to strengths of the order of $10^{14}$~G which correspond to a deformation $|\epsilon_B|\sim 10^{-8}$. 
The star is now stabilized with respect to r-modes up to frequencies
$\nu_{\rm cr}\approx 950$~Hz (see Eq.~(\ref{ncr})), but the maximum spin frequency is limited
by GW emission due to the magnetic deformation (see Eq.~(\ref{nmax1})).
The limiting frequency, for typical values of $\dot{M}\sim 10^{-9} M_\odot yr^{-1}$, corresponds to
$\nu_{\rm max}\sim 630 ~\dot{M}_{-9}^{1/5}~(\epsilon_B/10^{-8})^{-2/5}$~Hz.
\subsubsection{Evolutionary Path ($\bar{B}_{\phi,ort}\approx \bar{B}_{\phi,min}$)}
The second evolutionary path takes place if the condition $t \gtrsim \tau_{\rm v}$ 
is reached when $\bar{B}_{\rm \phi,ort}\approx\bar{B}_{\phi, {\rm min}}$.
After the rotation of the magnetic axis, the GW emission term $F^{\rm B}_{\rm g}$ becomes effective, 
with $|\epsilon_{\rm B}| \approx \epsilon_{\rm B,min}$. 
The new magnetic field will have poloidal and toroidal components of comparable
strength which stabilize the star against
$r$-modes up to frequencies $\nu_{\rm cr}\approx 500$~Hz.
Up to this frequency the magnetic deformation will thus be the only
effective cause of GW emission.
\begin{figure}
\begin{center}
\includegraphics[height=6cm, width=9cm]{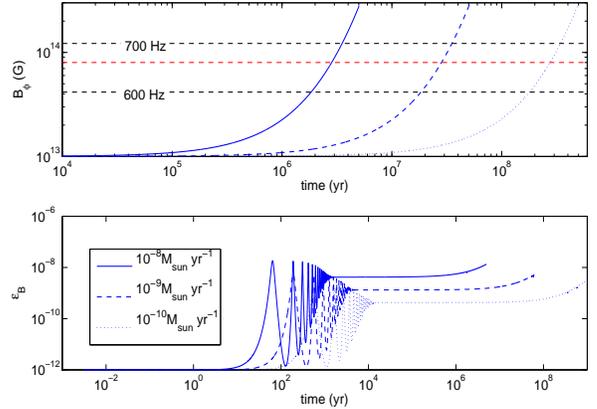}
\end{center}
\caption{\label{fig4}
\textit{Upper panel}: Temporal evolution of the generated toroidal magnetic field
when the star enters again the r-mode instability window at frequencies $\nu_{cr}\approx 500$~Hz
after the flip of the internal magnetic field ($B_{\phi, {\rm ort}}\approx B_{\phi, {\rm min}}\approx 10^{13}$~G).
We show also the spin frequencies of the star (dashed black lines)
obtained by Eq.~(\ref{ncr}). The red dashed 
line indicates the minimal strength of the generated toroidal magnetic field to have a new instability
by magnetic deformation (Eq.~(\ref{eq.Bcr})).
\textit{Bottom panel}: Temporal evolution of the r-mode amplitude $\alpha$ after the flip
of the internal magnetic field. We consider three values of mass accretion rate 
$\dot{M}=(10^{-8},10^{-9},10^{-10})\,\mbox{M}_\odot\,\mbox{yr}^{-1}$.}
\end{figure}
Mass accretion will continue to spin up the star, which may
eventually enter again the r-mode instability window thus starting the generation of new
azimuthal field. In Fig.~\ref{fig4} we show the temporal evolution
of the generated toroidal field and of the r-mode amplitude $\alpha$ 
(obtained by solving Eqs.~(\ref{eq6},\ref{eq7}))
when the star enters again the instability window at $\nu_{\rm s}\approx 500$~Hz.
The magnetic fields can grow further and the star may be subject to
a new instability when the magnetic field exceeds a value of the order of $10^{14}$~G
at frequencies $\nu_{\rm s}\approx 650$~Hz (see Eq.~(\ref{eq.Bcr}) and Fig.~\ref{fig4}).
\begin{figure}
\begin{center}
\includegraphics[width=9cm,height=6cm]{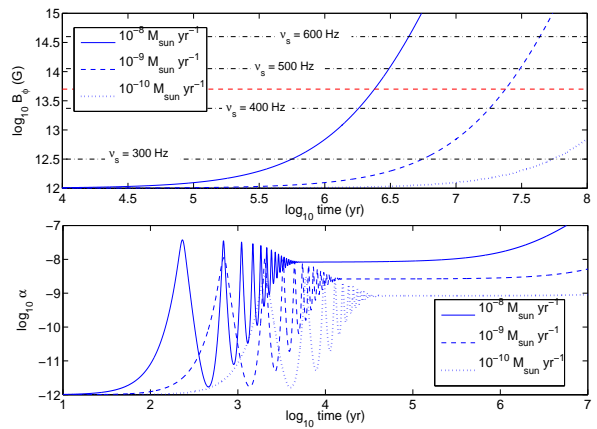}
\end{center}
\caption{\label{fig5} \textit{Upper panel}: Temporal evolution of the generated toroidal magnetic field
after the development of the Tayler instability.
Here we assume $B_d=10^{12}$~G. We show also the spin frequencies of the star (dashed-dotted lines)
obtained by Eq.~(\ref{ncr}). The red dashed line indicates the 
minimal strength of the generated toroidal magnetic field to have a new instability
by magnetic deformation (Eq.~(\ref{eq.Bcr})).
\textit{Bottom panel}: Temporal evolution of the r-mode amplitude $\alpha$ after the development
of the Tayler instability. We consider three values of mass accretion rate 
$\dot{M}=(10^{-8},10^{-9},10^{-10})\,\mbox{M}_\odot\,\mbox{yr}^{-1}$.}
\end{figure}
After the second \textit{flip} the evolution becomes 
similar to that described in the first scenario being again $B_{\phi, {\rm ort}}\approx10^{14}$~G
and $|\epsilon_B|\approx 10^{-8}$.
\subsubsection*{Tayler Instability}

The issue of the stability of the magnetic field generated by
$r$-modes still needs to be addressed.  
In the stably stratified environment of a stellar interior the Tayler
instability (or pinch-type) is driven by the energy of the toroidal magnetic field.
In a slowly rotating NS the instability is expected to set in at field strength 
$B_{\phi,\rm cr}\gtrsim 10^{12}$~G \cite{Spruit:1999cc,Cuofano:2009yg},
with unstable modes growing approximately on the Alfv\'en time--scale
as long as the ratio of the magnetic energy $E_{EM}$ to rotational kinetic energy $T_{rot}$
\begin{eqnarray}
\label{tyler-ratio}
 E_{EM}/T_{rot} &\sim& \frac{B^2R^3/3}{I\Omega^2/2} \nonumber \\
&\approx& 2\times10^{-8} \left(\frac{B}{10^{13}
   \,\mbox{G}}\right)^2\left(\frac{{\rm P}}{5 \, \mbox{ms}} \right)^{2}
\end{eqnarray}
is greater than 0.2 \cite{Kiuchi:2008ss}.
\\ 
In systems of interest to us, however, this ratio is expected to be
extremely small, as shown by the estimate of Eq.~(\ref{tyler-ratio}). In
this condition it is not clear at all that the Tayler instability will ever set in
~\cite{Kiuchi:2008ss,Lander:2010br,Braithwaite:2008aw}. It is
interesting to briefly
sketch the implications of this instability for the evolution of the
magnetic fields in fast accreting compact stars.
\\
The most important property of the Tayler instability in this context is that, 
after it develops, the toroidal component of the field produces, 
as a result of its decay, a new poloidal component which can itself be wound up, closing the
dynamo loop~\cite{Braithwaite:2005pa}. When the differential rotation stops,
the field can evolve into a stable configuration of a mixed poloidal-toroidal
twisted-torus shape, with the two components having a comparable
strength~\cite{Reisenegger:2008yk,Braithwaite:2005md,Braithwaite:2005ps,Braithwaite:2005xi}.
\\
Here we assume that the Tayler instability develops in the
Alfv\'en time--scale when $\bar{B}_{\phi}\sim 10^{12}$~G. 
At that point a poloidal field is generated, having a comparable strength,
and the star is stable against $r$-modes only for frequencies slightly larger than 200 Hz.
Mass accretion accelerates the star back into the instability region and a new
toroidal field can then develop. The growth of the modes, and the generation
of the new toroidal field, will start at a much higher initial poloidal
field, say $B_d \approx 10^{12}$ G.
\\
In Fig.~\ref{fig5} we show the
evolution of the r-mode amplitude and of the magnetic field in this scenario,
after the star re-enters the instability window. 
\\
The magnetic fields can grow further and the star may be subject to
magnetic instability when $B_{\phi}\gtrsim 10^{14}$~G at frequencies 
$\nu_{\rm s}\approx 450$~Hz (see Eq.~(\ref{eq.Bcr}) and Fig.~\ref{fig5}).
After the \textit{flip} of the internal magnetic field the evolution becomes once again
similar to that described in the first scenario with $B_{\phi, {\rm ort}}\approx10^{14}$~G
and $|\epsilon_B|\approx 10^{-8}$.

\subsection{Superconducting layer}
\label{supercond}
Protons in a NS core are expected to undergo a transition to a (type-II)
superconducting state at temperatures $\lesssim 10^9$ K~\cite{Baym:1969}.
Recent calculations of the proton energy gap indicate that this transition
should occur only in a limited density range in the outer
core~\cite{Baldo:2007,Akgun:2007ph}, 
corresponding to a spherical shell of thickness 
$\ell_{\rm s} = [1\mbox{--}3]$~km \cite{Baldo:2007,Cuofano:2009yg}. 
Accordingly, we assume protons in the inner core, a sphere of radius $R_1
= R - \ell_{\rm s}$, to be in a normal (as opposed to superconducting) phase. 

In a superconductor, components of the Maxwell stress tensor are enhanced by
the ratio $H_{\rm c1}/ B_{\rm s}$, with respect to a normal conductor~\cite{Eass:1977}. 
Here $H_{\rm c1} \approx 10^{15}$ G is the lower critical field~\footnote{This
is the minimal magnetic field strength required to force a non-zero magnetic
flux through a type-II superconductor.} and $B_{\rm s}$ is the
magnetic induction within the superconductor, which is always $\ll H_{\rm c1}$
in a NS~\footnote{Strictly speaking, the NS core should be in the Meissner
state with total flux expulsion. However, the large electrical conductivity of
the core material forces it into a metastable type-II state,
cfr.~\cite{Baym:1969}}. Several authors thus suggested~\cite{Jones:1975,Cutler:2002nw} 
that the deformation caused by a typical magnetic 
field  $\sim 10^{12}~{\rm G}$ would be $\sim 10^3$ times 
larger than given by, e.g. Eq.~(\ref{eccB}). This would have very important
implications for the scenario proposed here.

For a phase transition in a pre-existing magnetic field, Ref.~\cite{Akgun:2007ph}
argued that the magnetic induction in the superconducting outer core would be
reduced with respect to that in the inner normal core, as required to maintain 
stress balance at the interface between the two regions. As a consequence, no
significant change in the total NS ellipticity should be expected in this 
case.

In the context of the present work, the toroidal magnetic field is generated 
\textit{in the superconducting layer} at the expenses of the $r$-mode energy. 
We thus expect that a larger deformation of the NS can be obtained only if a
correspondingly larger amount of energy can be transferred from the modes to the toroidal field. 
Conversely, for a fixed energy transfer rate, one would expect a
\textit{weaker} field to be produced in the superconductor, such that 
the total magnetic stress - thus the induced deformation - would be the
same as for a normal conductor. 

A solution to the $r$-mode equations in an
inhomogeneous star would be required to solve this problem
self-consistently~\cite{Cuofano:2009yg}, which is beyond our scope here. However, 
we can gain some insight by considering the energy budget of the system.  

We adopt the expression of the magnetic damping rate in the normal and
superconducting case, following~\cite{Rezzolla:2001dh}
(cfr. Sec.~\ref{sezione3}). For the normal fluid in the inner core, 
the rate of production of magnetic energy is (cfr. Eq.~(\ref{Fmi}))
\begin{eqnarray}
\label{eq.magnetic-rate-normal}
\frac{{\rm d}}{{\rm dt}} E^{({\rm N})}_{\rm M} &=& \frac{8}{45 \pi} R_1
\Phi_{\rm B}^{\rm N}~\bar{B}^{({\rm N})}_{\phi} (t) \frac{\Lambda^{'}}{\Lambda} \,\alpha^2_{\rm N} (t)
\,\Omega(t) ~.
\end{eqnarray}
Here $\Lambda^{'}$ and $\Lambda$ are angular integrals of order unity defined in Ref.~\cite{Rezzolla:2001dh},
$\Phi_{\rm B,0}$ is the initial magnetic flux threading the inner core and 
$\bar{B}^{({\rm N})}_{\phi} (t)$ is obtained by taking the volume average
of the generated azimuthal field (cfr. Eq.~(\ref{eq2M})).

The superconducting layer has a volume 
$V^{({\rm SC})} = 4 \pi R^2_1 \ell_{\rm s}$ and we can write 
\begin{eqnarray}
\label{eq.magnetic-rate-typeII}
\frac{{\rm d}}{{\rm dt}} E^{({\rm SC})}_{\rm M} & = & \frac{H_{\rm c1}}{4 \pi}
V^{({\rm SC})} \frac{\Phi_{\rm B}^{SC}}{8\pi R^2_1} \frac{{\rm d}}{{\rm dt}} \Delta \tilde{x}^{\phi} \nonumber \\
 & \approx & \frac{H_{\rm c1} \Phi_{\rm B}^{SC}}{12 \pi^2} \ell_{\rm s}
\Lambda \alpha^2_{\rm SC} (t) \Omega(t) \, .
\end{eqnarray} 
As the $r$-mode energy is tapped by the newly generated azimuthal field, the 
magnetic energy density and Maxwell stress grow both in the inner and outer 
core. We thus expect that stress balance at the interface between the two 
regions will play a role in the subsequent development of  
$r$-modes. 
To obtain the rates of magnetic energy density production
$\dot{{\cal U}}_{\rm B}$ in either region we divide 
Eqs.~(\ref{eq.magnetic-rate-normal},\ref{eq.magnetic-rate-typeII}) by the 
corresponding volumes,  $V^{({\rm N})}$ and $V^{({\rm SC})}$. Hence 
\begin{equation}
\label{toroidal-fields-balance}
\frac{\dot{{\cal U}}^{({\rm N})}_{\rm B}}{\dot{{\cal U}}^{({\rm SC})}_{\rm B}}
=\frac{\bar{B}^{({\rm N})}_{\phi} (t)}{H_{\rm c1}} \frac{dB^{({\rm N})}_{\phi}}{dB^{({\rm SC})}_{\phi}}
\simeq \frac{\bar{B}^{({\rm N})}_{\phi} (t)}{H_{\rm c1}} \left[\frac{\alpha_{\rm
      N}(t)}{\alpha_{\rm SC}(t)}\right]^2 ~,
\end{equation}
having neglected numerical factors of order unity. 
If Maxwell stresses in the two zones need to meet an equilibrium condition, 
the above ratio should be of order unity. 
Accordingly, a stable evolution of the internal magnetic field requires that
$\alpha_{\rm SC}/\alpha_{\rm N}\sim \,[\bar{B}^{({\rm N})}_{\phi}(t)/H_{\rm c1}]^{1/2}$.
This relation can be considered as a boundary condition on the amplitude
at the surface separating the normal and the superconducting region.
If this condition on the $r$-mode amplitudes is not satisfied, we expect 
that the magnetic instabilities will affect 
the growth of $\alpha$ in the two zones acting to restore the stress balance.
From this, we conclude that, if the condition of stress balance were to
hold (at least in a time-average sense), azimuthal fields of different
strengths would be generated in the two zones (see Eq.~(\ref{toroidal-fields-balance})), namely
\begin{equation}
\label{azimuthal-fields}
B^{({\rm SC})}_{\phi} (t) \simeq \bar{B}^{({\rm N})}_{\phi} (t)
\frac{\bar{B}^{({\rm N})}_{\phi} (t)}{H_{\rm c1}}
\end{equation}
This implies that the total NS deformation is expected to be of the same order
of the deformation in a normal conductor. 
\\
Note that superconducting protons may coexist with superfluid 
neutrons, in a fraction of the neutron star core.
Pinning of neutron vortices to magnetic flux tubes may affect in this 
case both the growth of the magnetic field in the superconducting shell 
and the spin frequency evolution \cite{Alford2008PhRvB,Glampe2011MNRAS}.
However, this complicated interplay and its ultimate implications are still subject 
of much debate \cite{Srinivasan1990Sci,Ruderman1998ApJ,Glampe2011ApJ} 
and the problem is far from settled. The coexistence 
of the two mutually interacting superfluid and superconducting particles 
of different species is still controversial, and notable 
observational limits on it have been proposed \cite{Link:2003}. 
A detailed analysis of these effects and their inclusion in our numerical 
calculations are beyond the scope of this paper.

\subsection*{Discussion}

\begin{figure}
\begin{center}
\includegraphics[height=6cm, width=9cm]{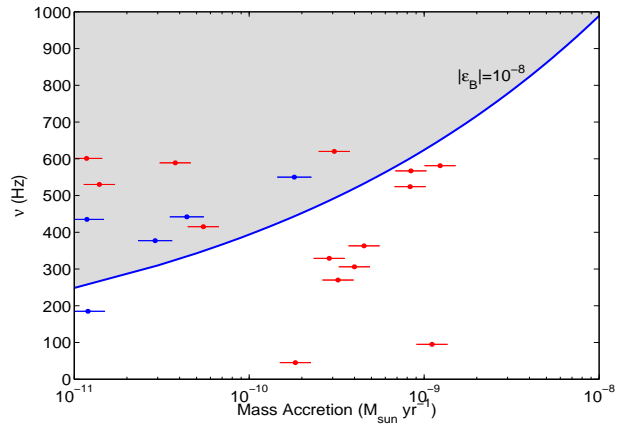}
\end{center}
\caption{\label{fig6} 
Limits on the maximum spin frequencies of accreting neutron stars 
with an ellipticity $|\epsilon_B|\sim 10^{-8}$ due to the magnetic deformation (solid line).
We show also the fastest accreting millisecond pulsars (blue points) and
the burst oscillation sources (red points)~\cite{Watts2008MNRAS}.
The stars can enter horizontally in the gray area when the mass accretion rate decreases
down to values $\dot{M}< 10^{-9}M_{\odot}\mbox{yr}^{-1}$ (see Discussion in Sec.~\ref{evol}).
}
\end{figure}

The analysis of the previous scenarios allows us to draw some general conclusion about the internal magnetic fields 
and the maximum spin frequencies of the fastest accreting NSs.
These conclusions are quite independent of the particular evolutionary path of the \textit{recycle}d stars.
In the cases where $B_{\rm \phi,ort}\ll 10^{14}$~G and $\epsilon_B\approx \epsilon_{\rm min}$,
the GW emission due to the magnetic 
deformation does not limit the spin frequency of the accreting stars that can re-enter the r-mode
instability window and generate new azimuthal field. 
The new toroidal component can reach strengths $\bar{B_\phi}\gtrsim 10^{14}$~G causing a new magnetic instability.
On the other hand, in the cases where $B_{\rm \phi,ort}\gg 10^{14}$~G 
the magnetic field rapidly decays by ambipolar diffusion to strengths again of order $10^{14}$~G.
It is clear that an equilibrium configuration requires a value of the internal magnetic field $B_{\rm \phi,ort}\approx 10^{14}$~G with an $|\epsilon_B|\approx 10^{-8}$. This internal configuration limits the spin frequencies at values  
$\nu_{\rm max}\sim 630 ~\dot{M}_{-9}^{1/5}~(\epsilon_B/10^{-8})^{-2/5}$~Hz preventing the star from further
increasing their magnetic field by re-entering the r-mode instability region. 
In Fig.~\ref{fig1} and Fig.~\ref{fig6} we show 
the limits on the maximum spin frequencies of accreting neutron stars as a function
of temperature and of mass accretion rate respectively.
In Fig.~\ref{fig6} we plot also the observational data of accreting millisecond pulsars
and of burst oscillation sources. The fastest spin frequencies ($\nu\gtrsim 650$~Hz) can be reached only for
mass accretion rates $\dot{M}\gtrsim 10^{-9}\, \rm M_{\odot} \rm yr^{-1}$ with a
spin up time-scale in the range $\approx [10^7\mbox{--}10^8]$~yrs.
Note that during this period of time, the internal magnetic field 
$B_{\rm \phi,ort}$ is not subject to significant decay.

Taking into account the maximum values of $\dot{M}$ for the NSs in LMXBs 
($\approx 10^{-9}\rm \,M_{\odot}\,\rm yr^{-1}$) we can estimate a maximum
spin frequency $\nu_{max}\lesssim 800$~Hz which is in agreement with the estimate $\nu_{max}\sim 730$~Hz 
of Refs.~\cite{Chakrabarty:2003kt,Chakrabarty:2004tp}. 

Note that the star can move \textit{horizontally} 
in Fig.~\ref{fig6} entering the \textit{forbidden} region (gray area) when the mass 
accretion rate decreases down to values $\dot{M}< 10^{-9}M_{\odot}\mbox{yr}^{-1}$.
An accreting spinning up neutron star is expected to move upwards in the 
white area in Fig.~\ref{fig6}. When its spin frequency reaches the limiting value 
indicated by the solid line the star cannot spin up further and its 
position remains close to the limiting line. If subsequently the mass 
accretion diminishes, the star drift horizontally to the left Fig.~\ref{fig6}. 
Therefore, in this interpretation, stars lying in the gray area of Fig.~\ref{fig6}
are presently accreting at significant lower rate than in the past.

\section{Detectability of the gravitational-waves emitted}

In this section we discuss the detectability of the gravitational radiation
emitted by \textit{recycled} millisecond NSs due their magnetically-induced distortion.
We consider a typical distance $d=10$~kpc of accreting NSs in our Galaxy and we calculate the average 
gravitational wave amplitude $h$.
\\
The instantaneous signal strain can be expressed as \cite{Stella:2005yz}:
\begin{equation}
 h \sim 1.5\times 10^{-29}\,\,k_\epsilon \,\,d_{10}^{-1} \,\, B_{\phi,13}^2 \,\, \nu_{500}^2~,
\end{equation}
where $B_{\phi,13}=B_{\phi}/(10^{13}\mbox{~G})$, $d_{10}=d/(10\,\mbox{kpc})$ and $\nu_{500}=\nu/(500\,\mbox{Hz})$.
\\
The spin frequency of accreting neutron stars changes significantly over a (long) timescale
\begin{equation}
 \tau_{sv} \equiv \frac{\Omega}{\dot{\Omega}}\simeq \left[ K_{m} \frac{\dot{M}}{\Omega} - {F_g^B}\right]^{-1} 
\gtrsim 10^6  ~\mbox{yr}~,
\end{equation}
where $K_m=(G/MR^3)^{1/2}/\tilde{I}$. 
Hence we can integrate for long periods $T_{obs}$. The minimal detectable signal 
amplitude $h_0$ is~\cite{Watts2008MNRAS}
\begin{eqnarray}
 h_0\approx11.4\sqrt{\frac{S_n}{T_{obs}}}
\end{eqnarray}
where $S_n$ is the power spectral density of the detector noise.
\begin{figure}
\begin{center}
\includegraphics[width=9cm,height=5cm]{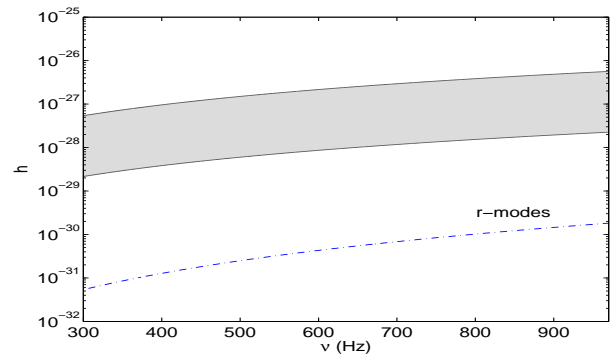}
\end{center}
\caption{\label{fig7} Averaged gravitational wave amplitude $h$ vs.
the spin frequency of the star $\nu$. The gray area is obtained assuming that
the accreting neutron star is located at a distance $d=10$~kpc and that
the internal magnetic field is in the range $\bar{B}_{\phi}\approx[0.2\mbox{--}1]\times 10^{14}$~G
and $k_\epsilon = 1$.
For comparison we show also the maximum amplitude of the GWs emitted by the star due to r-modes (dotted-dashed line)
after the rotation of the internal magnetic field or the development of the Tayler instability. 
We assume that $\alpha\simeq 10^{-8}$ (see Figs.~\ref{fig4} and \ref{fig5}).}
\end{figure}

\begin{figure}
\begin{center}
\includegraphics[width=9cm,height=5cm]{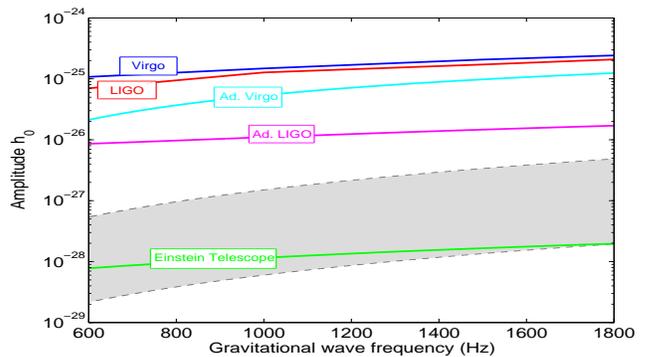}
\end{center}
\caption{\label{fig8} We compare the predicted amplitude (gray area) and the detectable
amplitude for LIGO (2010)~\cite{ligo}, Virgo~\cite{virgo}, Advanced LIGO~\cite{AdLigo}, 
Advanced Virgo~\cite{AdVirgo} and the Einstein Telescope~\cite{et} as a function of the frequency 
of the gravitational waves. Here we consider an integration time $T_{obs}=1$~yr. 
As in Fig.~\ref{fig7}, $d=10$~kpc, $\bar{B}_{\phi}\approx[0.2\mbox{--}1]\times 10^{14}$~G and $k_\epsilon = 1$.}
\end{figure}

In Fig.~\ref{fig7} we show $h$ as a function of the spin frequencies $\nu$ of the star,
while in Fig.~\ref{fig8} we compare the predicted and detectable amplitudes
as a function of the frequency $f=2\,\nu$ of the signal.

We show also the sensitivity curve of LIGO and Virgo and of the next
generation of detectors (Advanced LIGO, Advanced Virgo and Einstein Telescope). 

\section{Conclusions}

We have shown that accreting millisecond NSs can be deformed significantly by the very large magnetic 
fields generated by r-modes. A secular instability takes place when the magnetic distortion 
dominates over the maximal oblateness that the NS crust can sustain ($B_{\phi}\gtrsim 10^{13}$~G).
As a consequence of this instability, the angular momentum $J_*$ and the magnetic axis become orthogonal
on a timescale $\tau_{\rm v}$ given by Eq.~(\ref{eq.Tdis}) and the star begins to emit GWs.

We have shown that the GWs emission due to the magnetic deformation can limit the spin frequencies 
of accreting NSs. In particular we obtain a maximum spin frequency $\nu_{max}\lesssim 800$~Hz.
Our results are quite independent of the exact evolution of the internal 
magnetic field, that is difficult to estimate and depends on several unknown parameters, 
e.g. the growth timescale of the Tayler instability and the parameter $n$ that is related
to the \textit{flip} time of the generated toroidal field. 

In the end we analyzed the GWs emission due to the magnetic distortion.
Our results suggest that with an integration time of $1$~year, the next generation of detectors
(e.g. the Einstein Telescope) should be able to detect GW signals by accreting millisecond 
NSs located in our Galaxy.

\bibliography{bibliografia}

\end{document}